\input{aipcheck}
\documentclass[
    ,final            % use final for the camera ready runs
  ]
  {aipproc}
\layoutstyle{6x9}
\usepackage{mathtext,amssymb,amsmath,graphicx}
\begin{document}

\title{Horizons of Strong Field Physics}



\author{Johann Rafelski}
{address={Department of Physics, The University of Arizona, Tucson, AZ 85721, USA\\
and\\ Department f\"ur Physik der Ludwig-Maximilians-Universit\"at M\"unchen und
Maier-Leibnitz-Laboratorium, Am Coulombwall 1, 85748 Garching, Germany\\\phantom{line} }
}

\author{Lance Labun}
{address={Department of Physics, The University of Arizona, Tucson, AZ 85721, USA\\
and\\ Department f\"ur Physik der Ludwig-Maximilians-Universit\"at M\"unchen und
Maier-Leibnitz-Laboratorium, Am Coulombwall 1, 85748 Garching, Germany\\\phantom{line} }
}
 
\author{Yaron Hadad} 
{address={Department of Mathematics, The University of Arizona, Tucson, AZ 85721, USA\\
sand\\ Department f\"ur Physik der Ludwig-Maximilians-Universit\"at M\"unchen und
Maier-Leibnitz-Laboratorium, Am Coulombwall 1, 85748 Garching, Germany}
}

\begin{abstract}
Discussing the limitations on the validity of  classical 
electrodynamics, we show that present day laser pulse technology applied to
 head-on-collisions with relativistic electrons generates fields strong enough 
to permit  experimentation at the limits of validity of 
the Lorentz force, and the development of  experimental tests of Mach's principle.
We also discuss more distant opportunities for exploring the nature of 
laws of physics and the vacuum structure.
We then conclude that the predictions of quantum electrodynamics in the presence of critical
fields   are   not completely satisfactory and 
argue that  the study of Laser materialization into particle 
pairs opens a new domain   of quantum electrodynamics.
\end{abstract}

\maketitle
%03.50.De Classical electromagnetism, Maxwell equations (for applied classical electromagnetism, see 41.20.-q)
%04.80.Cc Experimental tests of gravitational theories 
%12.20.-m Quantum electrodynamics 
%12.20.Ds Specific calculations 
%52.27.Ny   Relativistic plasmas (refers to QED)
%52.38.Ph X-ray, gamma-ray, and particle generation
%52.38-r Laser-plasma interactions-in plasma physics
%-------------------
%52.40.Db  Electromagnetic (nonlaser) radiation interactions with plasma
%41.60.-m   Radiation by moving charges
%24.10.Pa Thermal and statistical models (refers to nuclear physics)
%13.60.Le   Meson production
%33.20.Xx Spectra induced by strong-field or attosecond laser irradiation
%                   (see also 33.60.+q Photoelectron spectra)
%52.25.Dg Plasma kinetic equations 
%52.25.Fi Transport properties 
%%%%%%%%%%%%%%%%%%%%%%%%%%%%%%%%%%%%%%%%%%%%
%% MAINMATTER
%%%%%%%%%%%%%%%%%%%%%%%%%%%%%%%%%%%%%%%%%%%%

\section{Critical Acceleration and Mach's principle}
%%%%%%%%%%%%%%%%%%%%%%%%%%%%%%%%%%%%%%%%%%%%%%%%%%%%%%%%%%%%

Strong fields produce strong acceleration, and thus are a probe of 
inertia. It is widely agreed   that acceleration requires as reference a frame 
against which inertia is measured, and this is true as much in  
a Lorentz covariant theory as in nonrelativistic dynamics. 
More than 100 years  ago Ernst Mach pointed out 
the need to quantify the inertial force with reference to what we 
call `an  equivalence class comprising all inertial frames of reference'. 
Mach chose the background  of fixed stars, i.e. cosmologically, 
the Universe at large\footnote{% 
Could inertial forces thus depend on the contact of a body with 
the Universe at large and thus be subject to control? This is 
a question which fascinates public at large.
 The reader will find googling `Mach's principle' many 
millions of hits.  Similarly, there are many web notes, and 
bona-fide research papers with  `Mach's principle' in the title.}. 
Given that Mach connected inertia to stars afar, there is no lack of misunderstandings 
about the meaning of  Mach's principle, and thus inertial force, 
in context of Einstein's or Newton's 
gravity. These have been recently explained and the meaning of `Mach's Principle'
categorized~\cite{Bondi:1996md}.

In our context we see two concepts contained in Mach's statement that play a role. \\
\indent {\bf a)} The measured value of inertial mass depends on all mass in the Universe (Mach1).\\
\indent {\bf b)} Acceleration is measured with reference to a 
select universal (fixed star) reference frame (Mach3).\\ 
Here the numbering in parentheses follows  Bondi and Samuel~\cite{Bondi:1996md} 
who further present nine other positions one can take regarding Mach's principle. 
  A review  of the subject goes beyond scope of this report and our 
discussion remains focused on the two items:
case {\bf a)} is  addressed by modern quantum field theory, with 
the Higgs field filling the Universe and providing the scale 
of masses, probably the scale of all luminiferous matter, perhaps the scale of gravity, 
in a yet-to-be-understood way.  

The Brans-Dicke~\cite{Brans:1961sx} extension to Einstein's Gravity (GR) was 
formulated with this goal and in that sense
it includes Mach in the theory of Gravity even at 
the Newtonian level by creating a framework for computing the
gravitational constant $G$ as the amplitude of a 
scalar field having many properties similar to the Higgs field
but employing totally different scales.  Note that 
as long as $G$ remains fixed and is not a dynamical field, 
Newtonian gravity and any theory which reduces 
to Newtonian gravity include some non-Machian contents.
Hence the leading Newtonian component in  Einstein's geometric 
theory of gravity (GR) must also be non-Machian. However, if we 
agree that Einstein's theory is an effective theory in the sense
we describe below, with the dimensioned matter-gravity 
coupling constant $G$ to be computed from a more foundational 
approach, then we can have  Mach's principle fully implemented. 
This is not the objective of this work, but a point of view
which we keep in memory as we address the shortcomings 
of the current understanding. 

Direct contact to light pulse high acceleration physics is made when considering 
the second  conceptual statement {\bf b)} contained in Mach's principle.
Today, we can study  
laser-electron interactions involving ultra high accelerations
achieving in special situations values which rival 
those expected at the event horizon of black holes. 
Einstein's equivalence principle requires 
physics at high acceleration to be identical with physics
in a strong gravitational field.  Thus in ultra high acceleration
experiments  we are addressing 
the understanding of gravity, and  more importantly, 
of inertia. 

An acceleration  of unit strength measured in 
natural units is achieved when the particle attains 
energy equivalent to its mass over the distance of its Compton wavelength.
For comparison, imagine that we accelerate an electron initially 
at rest over a distance of one meter to an energy of 1 MeV.  
In natural units, and in human experience units,   this amounts to an acceleration 
\begin{equation}
\dot v  = 7.54 \times 10^{-13}\,m_e  = 1.79 \times 10^{16}\,g
\end{equation}
We see that in natural units accelerators do not  accelerate much, 
yet expressed in terms of Earth's surface gravity, the scale is larger 
than one can imagine. 

Can one ever 
reach in laboratory `critical' acceleration limit, $\dot v\to 1 m_e$?  
The answer, amazingly, is yes, and it could be the next big experimental project.
Using an intense laser pulse for which the normalized 
vector potential is $a_0=eA_0/m_e$ we can generate a field to accelerate 
an electron of 100 MeV/micron by acting within the space of a quarter wavelength ($0.25\mu$) 
given $a_0=50$.  This corresponds to a gain in field strength 
of 8 orders of magnitude over the conventional accelerator
case we considered above.  Now relativity comes to help: 
if we look at the laser pulse from the frame of reference
 of a moving electron we gain a factor $\gamma(1+v^2)\simeq 2\gamma$.  
For exactly $\gamma=7 000$, that is an electron of 3.5 GeV, 
an observer riding on the electron experiences a   unit value of acceleration.

In order to achieve our goal of probing critical acceleration 
in a laboratory experiment other combinations 
of $a_0$ and $\gamma$ are possible. 
All it takes is placing an intense laser near an
accelerator or using a second intense laser pulse to form a relativistic 
electron beam.  In fact,
there has   been an experiment aiming to study strong field  effects
organized just in this way. The  SLAC  46.6 GeV  ($\gamma= 10^5$) electron 
beam was collided with most intense lasers available  13 years ago. 
Light pulses  were still 
much less intense, offering $a_0\simeq 0.5$  and as result the exploration
of strong fields occurred at well below critical acceleration limit ~\cite{SLAC}.

Today we can   reach the strong acceleration limit in the 
laboratory and explore experimentally the limits of our understanding of
inertia and Maxwell-Lorentz electromagnetism. 
The study of ultra high acceleration outside 
of the realm of GR   opens a new physics frontier
in that we deal with  `real' acceleration and inertial 
resistance to it. Acceleration is absent in the geometric 
general theory of relativity:
though we observe motion of a satellite as if there were 
a force, there is none; a satellite is free-falling.  
Einstein employed the equivalence principle to eliminate acceleration from 
his theory -- Newton's force arises from geodesic motion.

Since Mach3 is a statement about acceleration, 
Mach3 becomes irrelevant in the `classical' Newtonian limit of GR. 
A solution of the geodesic equation of motion for a probe particle 
of negligible mass in any gravitational field of `external'
character (a field unperturbed by the probe particle)
is by definition acceleration-free dynamics.
That is why a `good' theory such as Einstein's gravity, 
which is  Machian as much as it can be, reduces easily 
to non-Machian Newtonian gravity (non-Machian as long 
as $G$ is non-dynamical and not rooted in properties 
of space-time).

Only when we probe GR  beyond the Newtonian limit 
can the question  be posed: is Machian physics involved in GR? 
It helps to remember that a free fall is 
interrupted by the presence of matter.  A Machian 
effect arises already when we have dynamics of many 
bodies or one extended material body.  The best studied 
example is the rotation of the Earth dragging the nearby 
space-time manifold, leading to the Lense-Thirring 
effect~\cite{LT}.  The measurement of frame dragging amounts,
paraphrasing here Francis Everitt of Gravity Probe B~\cite{Everitt:2008zz}, 
to the measurement of a missing inch in the orbit of a satellite. 
The Gravity Probe B project 
refers the orientation of  a satellite to a fixed star
in order to  measure  that extra inch, 
thus directly  implementing Mach's suggestion to employ the fixed stars 
as the frame of reference. 

It is probably true that any effect in GR beyond the 
Newtonian theory is Machian, requiring for its evaluation 
reference to the space-time manifold on which matter exists. 
 Einstein pointed this out to Mach before completing 
his theory~\cite{EinstLetter}. Mach's Principle would
 seem to be addressed at this point. 
The issue remains that 
to  the best of our understanding, forces acting between, and 
on material particles, were not formulated in a form respecting a relation to 
 Mach3. Most vexing in our context is the fact that Maxwell-Lorentz 
electromagnetism is   non-Machian3. Paraphrasing Mach and Newton,
how can we  be sure that it is not the Universe that 
accelerates when we presume to measure acceleration? The only reason we 
can do physics is that in natural units the 
(electromagnetic) accelerations we encounter 
in normal life are negligible. 

On this basis, one could argue that Mach3 is not
satisfied by the fundamental forces we study in the microscopic world. 
Moreover, we know that these forces operate in the realm of quantum physics 
which is  inconsistent beyond the 
leading classical limit   with the one exceptional force 
(Gravity) that does seem to be consistent with Mach3.
However, the situation is much more complex.  
In  modern thinking, fundamental interactions are `effective', arising 
from the behavior and properties of the quantum vacuum state, the modern 
aether.  In some way we have not yet grasped, this means that just like 
with Brans-Dicke or Higgs approach, these interactions 
are  Machian, being a part of the modern aether theory.

%%%%%%%%%%%%%%%%%%%%%%%%%%%%%%%%%%%%%%%%%%%%%%%%%%%%%%%%%%%%%%%%%%%%%%%%%%%
\section{Aether, Vacuum,  Laws of Physics, and Matter}
Aether, the carrier of light waves, fell into disrepute
100 years ago due both to the absence of 
the effect of an aether drag in the Michelson-Morley experiment, and  
to Einstein taking the position in his 1905 papers that aether is 
unobservable. However, as is often the case, scientific positions 
evolve, and 15 years later Einstein wrote: 
\begin{quote}According to the general theory of 
relativity, space without aether is unthinkable; for in 
such space there not only would be no propagation of 
light, but also no possibility of existence for 
standards of space and time (measuring-rods and clocks), 
nor therefore any space-time intervals in the physical 
sense. But this aether may not be thought of as endowed with the
 quality characteristic of ponderable media, as consisting of parts
which may be tracked through time. The idea of motion may not  be
applied to it. [Concluding paragraph of: {\it Aether und 
die Relativitaetstheorie} (Berlin, 1920)]. 
\end{quote}
The last phrase is of particular relevance here and
we will soon return to the question in what sense  
aether is  a ponderable medium, with parts which can 
be tracked in time.

Einstein in effect postulated in 1920 the existence of 
a relativistically invariant aether. In this way he could 
connect  matter present  at any place in the Universe
with a common inertial frame of reference against  which acceleration
is measured. This creates a foundation for the implementation of Mach3.
Moreover, the aether is the carrier of physical qualities  and thus 
Einstein directly implemented Mach1.   

Our view of physics laws has evolved vastly since 1920 
and  the realization of Einstein's Machian  
objectives may be achievable today, partially because of the development
of new experimental tools discussed in this note.
The main new insight since Einstein's times is the recognition 
that with quantum mechanics and quantum field theory we acquire a 
structured vacuum state which has measurable physical properties. 
This vacuum structure arises from quantum fluctuations in the vacuum 
permitted by the uncertainty relation. 

To make these vacuum fluctuations
concrete in the early stages of development of quantum field theory, one 
imagined a network of points connected by ideal springs, and these 
could undergo oscillations which were the vacuum fluctuation modes. 
In the continuum limit of infinitely dense points, the 
transition from the quantum oscillator picture to a non-interacting 
quantum theory with particles occurs, and the zero-point oscillations 
of the oscillators become the (divergent) vacuum energy. 
In the numerical effort to solve
quantum field theory  on the lattice, we in fact return
to this quantum lattice of a three-dimensional 
chain of harmonic oscillators, but introduce a
gauge invariant action.

The energy of the quantum ground state diverges in two 
ways: since it is extensive, any finite energy density 
diverges with the volume size.  Moreover, when we allow 
the distance  of lattice points  to shrink, we allow
fluctuations of arbitrarily large momentum and thus
in the continuum limit we arrive at:
\begin{equation}\label{zero}
\frac{E}{V}=\frac{\pm g}{2}\int \frac{d^3p}{(2\pi)^3}\sqrt{m^2+p^2}
\end{equation}
where $g$ is the degeneracy, due to two spin states of fluctuations 
at same momentum, and also due to fluctuations of both particles and 
antiparticles -- thus for 
electrons, positrons $g_e=4$. The factor $1/2$ originates
in the zero point energy of each single harmonic oscillator, $E_0=\omega/2$.

The overall sign changes between Fermions (-), and Bosons (+).  
It should be remembered that for photons we have $g_\gamma=2$ since there 
are only two (transverse) degrees of freedom of free photons, and no 
antiparticles.  Thus, there is a partial cancellation of leading infinity 
from photons with that from electrons contributing with opposite sign,
and the dominant vacuum energy
remains divergent in quantum electrodynamics (QED).
In general, to cancel the zero point energy 
by symmetry,  the additional  `super'-symmetry is required, which must be 
badly broken in nature. Still, this symmetry gives birth to new particles, 
including candidates for dark matter. 

We can measure the energy in the vacuum against this infinite 
value. The best known example is the Casimir energy which arises  
between  two conductive plates. Since photon fluctuations have to 
end on the plates, fewer quantum fluctuations can exist 
between the plates, and the greater fluctuations 
outside press the plates together. Measurement of the Casimir effect
is routinely possible today. Because the effect requires presence of
matter, some argue that the Casimir effect does not require a change
in vacuum structure of the  fluctuations. While we do not share
this opinion, we note that a much simpler effect exists which 
directly relates to vacuum fluctuations, namely the vacuum polarization.

When we apply an external electromagnetic field to the vacuum, 
the fluctuations of electrons and positrons separate and we 
find that  the vacuum has a dielectric polarization property. 
In order to arrive at a finite observable value we need to 
redefine electron charge (charge renormalization).  This means that
the measured electrical charge, $e=\sqrt{\hbar c/137} $ arises
from a bare charge  $e_0\simeq \sqrt{\hbar c/40}$ which we would 
observe were we able to probe at  Planck length scale $L_P$,
the shortest physical length in our Universe, 
$L_P\equiv \sqrt{\hbar G/c^3}=1.61 \times 10^{-35}$m. Because the 
leading effect of the polarization is to alter the charge, 
in QED we observe polarization effects that strengthen the 
magnitude of the applied field. For example, the Coulomb potential near
to the atomic nucleus is stronger by a few parts in a thousand compared
to our expectations. 

While vacuum structure effects  in QED are relatively subtle
in normal laboratory environment, the situation is drastically 
different for quantum chromodynamics (QCD), the theory
of strong interactions between 
nucleons. The much larger color charge of gluons,
the `photons' of QCD, alters the nature of the vacuum state 
completely, in a way that makes it impenetrable to the motion 
of color charges of quarks and gluons and so generating quark confinement. 
This also implies that the vacuum must have a physical 
property related to this radical change in its 
structure. Indeed, the vacuum expectation value of 
the square of the gluon field fluctuations is non-zero, and we
give this effect the name `gluon condensate', a dimensioned 
quantity associated with every point in our Universe.  
We see that by the way of the quantum vacuum, 
Einstein's relativistically invariant aether is back. 

Long ago, at the beginning of time, 
when the Universe was much hotter, at a temperature
that exceeded 30 000 times that in the core of the sun,  
quarks and gluons could roam free. Since the quark-gluon
interaction charge, the color, is `ionized', the new 
state of matter is referred to  as the quark-gluon plasma or 
for short, quark matter. The expansion of the 
quark-gluon Universe cooled the plasma until the free color charges
were frozen into the hadrons we find in our Universe today. 
However, in each hadron containing quarks a `piece' of  
the non-confining `vacuum' is captured,  
allowing quarks to exist there. In colloquial language this is the quark-bag.  
In this way each proton, or neutron, is a carrier of a piece of 
the `wrong' aether from the early Universe. 

All matter particles of finite size therefore break 
Einstein's rule that specifically forbids tracking
vacuum pieces in time.However, this feat is achieved in a Lorentz invariant
way, since the matter particles are  characterized  by mass and  helicity,
(projection of spin onto axis of motion) which are
the two Casimir operators of the Poincare group of all 
space-time symmetry (translation, rotation) transformations. There are
further characteristic properties, such as charge, which 
imply that a group greater than the four dimensional Poincare group 
characterizes our space-time manifold.

We extend the concept of Einstein's relativistic aether by adding  matter and
allowing the coexistence of several forms of  aether captured within
matter particles, which
can be tracked and are ponderable. In this way the problem of 
finite size of massive matter particles is resolved and
the magnitude of their large mass understood as being due to 
the   zero point energy of more fundamental particles captured within. 
The first step of Pauli's `what is matter'
problem  has been resolved. The second step, the relationship of 
matter to fields, remains. 

Much of the above insight about QCD and vacuum participation 
in the explanation of structure of matter is a paradigm, a 
way of thinking about what we see in nature, supported
by elaborate  interpretation of experimental data. The 
example of the Geocentric vs. Heliocentric systems of celestial motion
shows that without more direct and drastic 
experimental evidence, quark confinement could, in principle, be differently
interpreted another day.  For this reason it is necessary to 
demonstrate directly the role of confinement in particle structure.

To achieve this we would like to recreate in a laboratory 
experiment the conditions of the early Universe, that is, the
other type of aether in which quarks roam freely. The 
formation of quark-gluon plasma seems possible in high energy heavy ion 
collisions. It is believed that we can smash the 
boundaries between the individual hadrons and  fuse  into 
one the quark content of individual nucleons.  In the reaction 
all  particles are heated tremendously and  
the confining structure of the vacuum that surrounds us today is melted.
An extended---albeit very small---domain of space is created for an 
exceedingly short time in which quarks
can roam freely as they did in the early Universe. 

The draw-back in this program of research is that the laboratory
micro-bang is indeed very small since the energy content we can 
deliver with particle accelerators 
is below 10 erg ($10^{-6}$\,J).  This energy restricts the 
size of QGP we form to nuclear length scales $R\simeq 6$\,fm. 
Since such a small drop of early Universe can expand unconstrained, 
with edges moving near velocity of light, its lifespan is bounded
to $3R/c\simeq 6\times 10^{-23}$\,s.  These relatively small values invite 
new efforts using light pulses, which are known to be much more
energetic than particle collisions. The challenge is here to learn
how to focus a good fraction of the MJ (megajoule) energy of a laser pulse 
into nuclear size.
Aside from present efforts to focus and/or compress light wavelength,
we can also explore multi-ion collisions once 
laser  intensity is at the level to directly accelerate heavy 
ions to relativistic energies.  Ion acceleration at this scale will be 
possible just above $a_0\simeq 5,000$, still two orders  of magnitude below 
the field strength  needed to reach critical acceleration. 
While this objective is as yet beyond today's technology horizon, 
it constitutes a worthy challenge for the future, with considerable pay-off in 
terms of study of the quantum vacuum, the modern  aether. 

An interesting element of the discussion presented is that  
elementary particle properties and
thus their interactions are subject to change.  In that
sense, they are not truly elementary.  The question which comes to mind is,
if elementary particles melt and change, can laws of physics
melt too?  Many if not all 
elementary interactions are effective interactions. A well-known QED 
example is light-light scattering, impossible in Maxwell theory but
present in the quantum vacuum due to the effective action of  Euler and 
Heisenberg, which is essentially an evaluation of Eq.\,(\ref{zero})
with vacuum fluctuation energies modified by the presence of the electromagnetic
field.  Further discussion of this effective action is offered below,
see Eqs.(\ref{Ef},\ref{Bf}).

Clearly, if  properties of the quantum aether generate new interactions, 
we can expect that the nature of the interactions we hold to be 
fundamental, and more generally all laws of physics, depends on the
nature, condition and type of the vacuum state. `Melting' the QCD vacuum 
on a relatively-speaking macroscopic scale using light pulses will 
help us to understand these questions.  In fact, there 
is a hierarchy of vacuum structure states, and beyond QCD we have the
Higgs vacuum, which could perhaps be melted if we were  able
to compress a 10 kJ light pulse into the volume of an elementary particle
such as a proton.
According to current thinking, in this state the Higgs
vacuum properties  would dissolve and the masses of 
all particles would go to zero or have the neutrino mass scale.
We are looking forward to
learning more about the topic in the context of the forthcoming study
of the Higgs particle at LHC. 

%%%%%%%%%%%%%%%%%%%%%%%%%%%%%%%%%%%%%%%%%%%%%%%%%%%%%%%%%%%%%%%%%%%%%%%%%%%%%%%%%%%%%%%%%%
\section{What is wrong with Electromagnetism}
After this grand tour of modern particle and field concepts we
are ready to  reconsider the theory of electromagnetic interactions
and to describe its shortcomings. The covariant form of the  Lorentz force is  
\begin{equation}\label{LF}
m\dot u^\mu=qF^{\mu \alpha}u^{\beta}g_{\alpha \beta} ,
   \qquad u^\mu=(\gamma,\: \gamma\, \vec v),
    \qquad  g_{\alpha \beta}= {\rm diag}(1,-1,-1,-1).
\end{equation}
The extension  of the Lorentz force   in the presence of strong acceleration 
has been a topic of intense research effort for the past 100 years, 
beginning as soon as the form of the force was written.  
The deficiency is easily seen: the Lorentz force does not 
`know' that the accelerated charged particle radiates. 

We all know 
of synchrotron radiation and compute it as students and set 
it as an exercise when we teach.  But few of us reached the 
last section of David Jackson's third edition of 
{\it Classical Electrodynamics} which reveals that 
the radiation emitted alters in principle the dynamics 
of the charged particle source.  A search in literature 
produces several distinct methods of accounting for this 
effect (see table \ref{tab:RRModels}). 
Each of these new force equations produces 
a different outcome and we recognize that all are not more than 
a patch  which is only meaningful when the acceleration 
is tiny.  For a recent comprehensive and rigorous first order 
study of the effect,  see~\cite{Gralla:2009md}.

%%%%%%%%%%%%%%%%%%%%%%%%%%%%%%%%%%%%%%%%%%%%%%%%%%%%%%%%%%%%%%%%%
\begin{table} [h!]
\caption{Models of  radiation-reaction extensions of the Lorentz force
 \label{TableRR}
}
\begin{tabular}{|c|c|}
\hline
LAD~\cite{Dirac:1938nz} & 
$\mathbf{m\dot{u}^\alpha=qF^{\alpha\beta}u_{\beta}}+m\tau_0 \left[\ddot{u}^\alpha
  -u^\beta \ddot{u}_\beta u^\alpha\right]$ \\ 
\hline
Landau-Lifshitz~\cite{LandauLifshitz} &
 $\mathbf{m\dot{u}^\alpha=qF^{\alpha\beta}u_{\beta}}+q\tau_0\left\{F^{\alpha\beta}_{,\gamma} u_\beta u^\gamma 
+ \frac{q}{m}\left[F^{\alpha\beta}F_{\beta\gamma}u^\gamma-(u_\gamma F^{\gamma\beta})(F_{\beta\delta}u^\delta)u^\alpha\right]\right\}$ \\ 
\hline
Caldirola~\cite{Caldirola:1979mi} & 
% $-\frac{m}{\tau} \left[u^\alpha _- - u^\alpha (u \cdot u_-) \right] = qF^{\alpha\beta} u_\beta$ \\ 
$\mathbf{0=qF^{\alpha\beta}\left(\tau\right)u_{\beta}\left(\tau\right)}
   +\frac{m}{2\tau_0}\left[u^{\alpha}\left(\tau-2\tau_{0}\right) 
    - u^{\alpha}\left(\tau\right)u_{\beta}\left(\tau\right)u^{\beta}\left(\tau-2\tau_{0}\right)\right]$ \\ 
\hline
Mo-Papas~\cite{Mo:1970vy} & 
 $\mathbf{m\dot{u}^\alpha=qF^{\alpha\beta}u_{\beta}}+q\tau_0 \left[F^{\alpha\beta}\dot{u}_\beta 
 + F^{\beta\gamma}\dot{u}_\beta u_\gamma u^\alpha\right]$ \\ 
\hline
Eliezer~\cite{Eliezer} &
 $\mathbf{m\dot{u}^\alpha=qF^{\alpha\beta}u_{\beta}}+q\tau_0\left[F^{\alpha\beta}_{,\gamma} u_\beta u^\gamma 
+ F^{\alpha\beta} \dot{u}_\beta - F^{\beta\gamma} u_\beta \dot{u}_\gamma u^\alpha\right]$ \\ 
\hline
Caldirola-Yaghjian~\cite{Yaghjian} &
 $\mathbf{m\dot{u}^\alpha=qF^{\alpha\beta}\left(\tau\right)u_{\beta}\left(\tau\right)}
   +\frac{m}{\tau_0}\left[u^{\alpha}\left(\tau-\tau_{0}\right) 
    - u^{\alpha}\left(\tau\right)u_{\beta}\left(\tau\right)u^{\beta}\left(\tau-\tau_{0}\right)\right]$ \\ 
\hline
\end{tabular}
\label{tab:RRModels}
\end{table}
%%%%%%%%%%%%%%%%%%%%%%%%%%%%%%%%%%%%%%%%%%%%%%%%%%%%%%%%%%%%%%%%%

Equations of motion are usually obtained by means of an action principle.
Thus it is important to note that the action principle of electromagnetism
does not implement at all the ability of accelerated charged particles to radiate.
The action comprises three elements: Maxwell field action, Matter-Field 
interaction (gauge invariant), and charged matter dynamics.  
These are written in the covariant form 
\begin{equation}\label{Action}
{\cal I} = -\frac 1 4 \int d^4x \:F^2 +q \int_{\rm path}dx \cdot A 
        + \frac {mc} 2 \int_{\rm path} d\tau (u^2-1).
\end{equation}
Two natural constants are introduced: `$q$' describing the relation
of matter to field  and `$m$' describing the inertia of matter. Many books
write  $mc\int d\tau $ for the last term and struggle with the 
constraint $u^2=1$, and some books   write  the middle 
term in the form $e \int_{\rm path}d\tau u \cdot A$ introducing 
explicitly the 4-velocity $u^\mu\equiv dx^\mu/d\tau$. 

The  two first terms in Eq.\,(\ref{Action}) 
 assure  that, upon variation with respect to the field, accelerated charges 
radiate according to the Maxwell equations with accelerated sources. 
The second and third term, when
varied with respect to the form of the material particle world line, 
produce the Lorentz force Eq.\,(\ref{LF}) 
(the first two terms in the   table \ref{TableRR}).  
One of probably several reasons the standard action fails is thus that
we add matter and specifically inertia in an ad hoc fashion to the action.  
Without doubt, inertia represented by the $m$-term, is from the theoretical 
point of view, the least satisfactory of the three terms in Eq.\,(\ref{Action}).
It is constrained merely by the nonrelativistic limit.  
The other two action terms are 
constrained also by gauge invariance. 
If, for example, mass were made out of field energy,
this would introduce a relation of fields and velocities,
helping to create radiation reaction terms in matter dynamics.  
In fact all studies of radiation reaction must address the 
renormalization challenges originating in the  electromagnetic energy component 
in the material mass.

In absence of a theoretical framework two different approaches 
have  been pursued: modifications of inertia or modification of field dynamics.\\
{\bf \phantom{I }I:} We can modify the Lorentz Force Eq.\,(\ref{LF})
  as is shown in  table \ref{tab:RRModels},
exploiting the known expression for the power
radiated ($P\propto \int a^2 dt$).  Considering 
further that the power radiated depends at least in principle on
the form of the generalized Lorentz Force  used to obtain the world lines of particles, 
there has been considerable freedom in introducing different modifications  
which only agree with each other at  first order in $\tau_0$.\\
\indent a) Given $P$ for radiation emitted, the unique result is  the 
Lorentz-Abraham-Dirac (LAD) equation~\cite{Dirac:1938nz}.
This equation has not been widely accepted, since among its solutions are  unphysical 
components which need to be eliminated using   the knowledge of 
the dynamics at an infinite future time.\\
\indent b)  The Landau-Lifshitz (LL) equation~\cite{LandauLifshitz}  is of particular interest
since this   generalization of the Lorentz Force is using velocities 
and fields, and for this reason does not introduce the LAD problems.
It has  gained  the status of a valid theory. However, 
absence of an action principle from which this, or any other similar form, can be derived  
renders these claims~\cite{Spohn:2004ik,Rohrlich} wanting, showing that these are also
ad-hoc modifications.\\
{\bf \phantom{I}II:} Since strong acceleration is related to gravity 
there were two efforts to arrive at unification of electromagnetic and 
gravitational field phenomena,
and a third effort simply introduces a limit to  acceleration
by limiting the physical field strength: \\
\indent a) Weyl's electromagnetism derives from the requirement 
that the metric vanishes under double covariant differentiation 
(gravitationally covariant and gauge covariant in EM theory). 
 For a better insight into  
reasons this theory fallen out of favor we recommend 
the excellent review of O'Raifeartaigh  and 
Straumann~\cite{O'Raifeartaigh:2000vm}. Weyl adapted his 
framework to   quantum theory and did not claim its 
relation to gravity later in his life. We use his gauge invariant 
derivative daily and yet few remember that Weyl introduced 
the concept of gauge invariance and covariant derivative to field theory. \\
\indent b) The Kaluza-Klein Theory remains a candidate for a  
unification of Gravity and Electromagnetism.
This theory has never been abandoned, for a recent 
review see~\cite{Overduin:1998pn,Salam:1981xd}. However, 
only once this field theory can be consistently complemented 
with  matter will it be suitable for our purpose.  For this 
reason and because it is the entry point to string theory, 
theorists marched into greater KK string dimensionality 
seeking the solution of what Pauli termed `the matter problem.' 
The Kaluza-Klein theory is still a theory without matter,
and  in a theory lacking entirely the material  electron, 
we cannot explore its response to acceleration.\\
\indent c) The Born-Infeld theory of electromagnetism~\cite{Born:1934gh} included 
a manifest limit to acceleration modeled after the limit to 
velocity seen in relativity: the Lorentz-force was bounded 
from above by an explicit limit to the field strength.  
This approach also allows to interpret the mass of 
an electron as entirely electromagnetic, however
this produces radiation effects that can be very large.
A more serious practical  problem with 
this approach is that one can see 
deviations from linear electromagnetism in precision 
study of high $Z$ atomic spectra, and the resulting 
high limiting limiting value found for the Born-Infeld
 field  in order  that this effect is invisible 
is at least 50 times greater than the 
 critical acceleration field~\cite{Rafelski:1971xw,Rafelski:1973fm}.

To close this general discussion we once more note that   
there is no action available for any
of the radiation reaction forces we show in table \ref{TableRR}. In general all 
patches of force equations not originating from an action violate energy 
conservation. Therefore  if energy
conservation is implemented this becomes a further patch of a patch, and 
we are losing  further control of the missing physics. 
The long development of the best known patches to Lorentz force, 
the LAD   effective radiation reaction force and of the 
related Landau-Lifshitz radiation reaction force are testimonials 
to the fact that combination of Maxwell equations with Lorentz force 
acting between charged particles, even if solved self-consistently, 
does not provide a full theoretical framework describing strongly 
radiating charged particles.

 This quite stunning insight is not 
new---one can see a trail of research reaching back to beginning 
of Lorentz and Maxwell theory.  It is also clear that perhaps 
as long as 50 years ago, as soon as lasers were invented, 
someone somewhere noticed that the best opportunity to experimentally
test  radiation reaction theory is to combine intense lasers with 
moving electrons. 50 years of laser technology development finally
allows the planning of experiments in this domain. 

 The trail of publications is thick, and 
thus we will present only the head of the trail, the latest 
reports at this meeting~\cite{Naumova, Keitel}.  This said, 
the  new element which we address in this paper
is that exploration of radiation reaction 
{\bf  is a research program on inertia, 
Mach's principle and strong gravity, the development of new 
fundamental action principle,  and not merely  a
measurement  of an obscure and not uniquely 
defined radiation  effect.}  

We stand before an enormous scientific opportunity. 

%%%%%%%%%%%%%%%%%%%%%%%%%%%%%%%%%%%%%%%%%%%%%%%%%%%%%%%%%%%%%%%%
\section{Experiments on Radiation Reaction}
In order to gain   quantitative insight  
about the experimental conditions needed 
for the study 
of inertia, we ask when the radiation reaction 
becomes important in the description of dynamics of 
charged massive particles.
For an electron traveling against an electromagnetic laser field, 
the radiation-reaction effects dominate the dynamics of 
the electron when\cite{Yaron}
\begin{equation} \label{RRDRCriterion}
(\omega \tau_0) \,\gamma_0 \,a_0 ^2 \sim 1
\end{equation}
where $\omega$ is the frequency of the laser wave, 
$\gamma_0$ is the initial $\gamma$-factor of the 
electron and %$\gamma_0=\frac{1}{1-(v_0/c)^2}$ and
\begin{equation} \label{Tau0}
\tau_0=\frac{2}{3}\frac{e^2}{mc^3}
\end{equation}
is a constant with dimensions of time, whose numerical 
value for the electron is $\tau_0=6.24\times 10^{-24}\,{\rm s}$. 
The origin of the radiation-reaction criterion Eq.\,(\ref{RRDRCriterion})
is recognized by inspecting the relative scale of Lorentz and 
 Landau-Lifshitz radiation correction force
in table \ref{TableRR}.

\begin{figure}[h!]
\includegraphics[width=.75\textwidth]{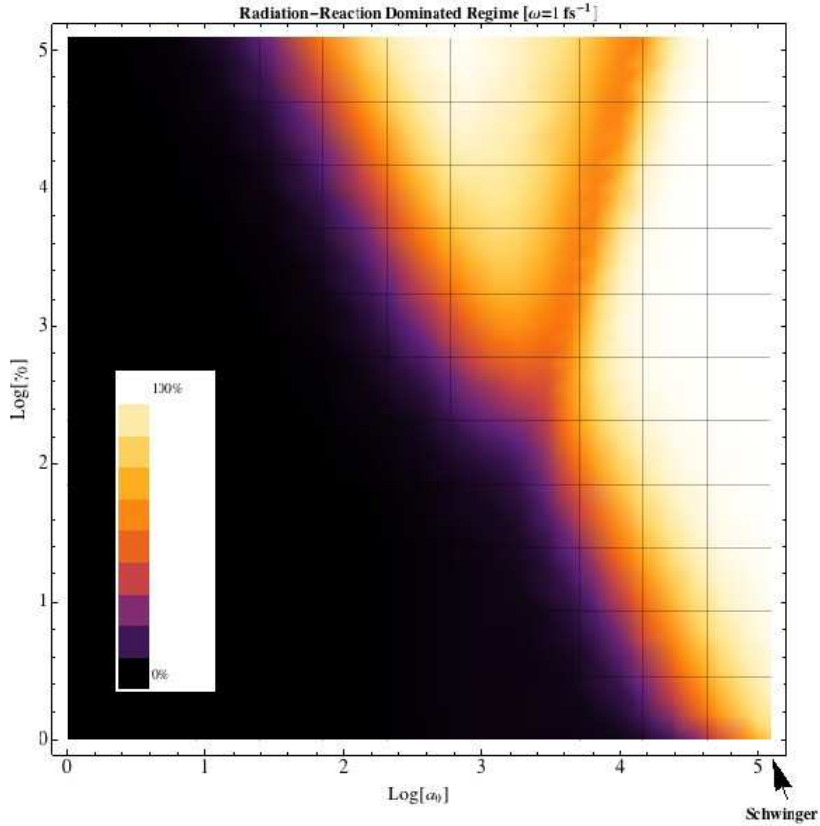}
\caption{\label{RRDRGraph} Demonstration of the criterion 
in Eq.\,(\ref{RRDRCriterion}) which determines in what 
conditions are the radiation-reaction effects important.  
The density presents the relative deviation in energy 
of a particle obeying the Landau-Lifshitz equation 
compared to a particle subject to only the Lorentz 
force.  One sees that above the critical line the 
radiation-reaction effects completely dominate 
the Lorentz dynamics of the particle. For an electron at
rest this requires   fields at the Schwinger limit 
($E\to 1\, m^2/e$ corresponding to $a_0\simeq 500\, 000$), but
the required field ($a_0$) drops rapidly as we exploit the 
large relativistic $\gamma$ factor.}
\end{figure}

This quantitative criterion is verified in figure \ref{RRDRGraph} 
where we show the relative deviation in the energy predicted by 
an equation of motion including radiation reaction 
(the Landau-Lifshitz equation) compared to the Lorentz force.

%%%%%%%%%%%%%%%%%%%%%%%%%%%%%%%%%%%%%%%%%%%%%%%%%%%%%%
%%%%%%%%%%%%%%%%%%%%%%%%%%%%%%%%%%%%%%%%%%%%%%%%%%%%%%%

The quantum field theory of charged particles, quantum electrodynamics, 
displays a similar pathology near to where acceleration turns 
to unity in presence of strong uniform fields: 
this  quantum critical field renders the vacuum state unstable 
to conversion into a gas of electron-positron pairs. We will review
this matter in more detail in next sections. This is only
an apparent limit to acceleration strength   since we 
can explore behavior of charged particles in collisions 
with non-uniform fields as discussed above.

The question rings loud at this point of our discussion:
is there   a radiation-reaction effect when we scatter
strongly interacting particles from each other?  Strong
interactions can impart large acceleration on electromagnetically 
charged particles and thus one would expect that there is 
a serious violation of conventional theoretical expectations
for production of radiation. In fact, in $pp$ and $\pi p$ 
collisions a  very serious photon 
excess   was identified in painstaking 
analysis by Martha  Spyropoulou-Stassinaki 
of Athens University~\cite{Belogianni:2002ib}. It has 
remained without explanation. 

In heavy ion collisions
excess of both photons and lepton (electron, muon) 
pairs have been seen and remained largely 
unexplained --   these radiation 
reaction effects have obscured the usefulness of photons and leptons
as signatures of quark-gluon plasma. However, because collisions involving
many nucleons (of type $A$) in nuclei  are much more difficult to interpret, 
the situation is  not as experimentally clean and 
clear as it is in the more elementary  $pp$ and $\pi p$ collisions.
Moreover, these results are often scaled in cascade programs to extract
the expected heavy ion backgrounds, and thus any measured $pp$
photon and lepton pair enhancement becomes part of the $AA$ background.
 
The story does not end here; the theory of interacting quarks and gluons,
quantum chromodynamics is patterned after quantum electrodynamics and
thus any shortcoming that one finds in QED will be present in QCD, especially
at high energies where the classical limit prevails. 
Moreover, since the quark-gluon coupling is as much as 50 times stronger,
in suitable circumstances the strong acceleration and radiation reaction 
effects could appear much  more easily in QCD. A  very pertinent effect
is the observation of 
the strong stopping power of quarks and gluons  in quark-gluon plasma, 
the effect of `jet quenching' \cite{quench}.
Theories can be  developed to explain this 
within the conventional  theoretical framework, yet
the fact remains that one must stretch all parameters and reaction mechanisms
in order to describe these effects. Thus it is safe to say that there 
would be no contradiction  with strong interaction physics 
if  a theoretical framework were to appear
in which radiation reaction of gluons and photons (both couple to quarks)
acts much stronger than inferred within small-acceleration Lorentz-type theory.

This short section and foregoing discussion show that there is some 
experimental and theoretical progress and an inkling presence 
of radiation-reaction effects. However, clearly a dedicated effort must be made
to understand critical acceleration or simply, inertia.  
As we have argued, within the  near future light 
pulse collisions with relativistic electrons will 
provide the experimental  opportunity for exploring this most 
challenging question from the experimental perspective.  

%%%%%%%%%%%%%%%%%%%%%%%%%%%%%%%%%%%%%%%%%%%%%%%%%%%%%%%%%%%%%%%%%%%%%%%%%%%%%%%%%%%%%%%%%%%%%%%%%%%
\section{Critical acceleration, QED and temperature}
The classic prediction of the Euler-Heisenberg-Schwinger 
analysis of QED in strong constant external fields 
(for a recent review see~\cite{Dunne:2004nc}) is that 
at the unit field strength, $E_s \equiv m_e^2/e=1$ 
in natural units, the electrical field becomes massively
 unstable, collapsing via materialization into electron-positron 
pairs on a microscopic time scale~\cite{Labun:2008re}.   
This is a fascinating result and the measurement of 
production of matter from fields in vacuum by ultra 
intense laser fields is the signature experiment 
in strong field physics~\cite{Mourou:2006zz,Marklund:2006my,Ruffini:2009hg}.

Though pair production goes under the premise of a 
test of QED in strong fields, nobody ever questions the theoretical framework
when the fields become strong.  The general assumption is that QED 
is correct, and thus exploration of the pair production 
mechanism is the mainstream of  the current study.  
The problem is that we already know that the 
classical theory of electromagnetism is not 
complete, as has been remarked in preceding sections. 
Can the quantum field theory built upon it be complete?  
This is the question that the fields-to-matter 
experiment can  answer.

In our opinion, QED at the critical field strength cannot 
be complete, and it is very likely that before 
laser fields strong enough to break the vacuum are
created, theoretical clarity will be reached on this 
issue. In order to see what is the problem we inspect  
the effective action of Euler and Heisenberg, which 
has the privilege of being one of a few exact results
in this area of physics. Indeed,  one does not need to look far to see that 
something is missing in this expression.  

We use here the connection between 
Euler-Heisenberg-Schwinger (EHS) 
action and temperature~\cite{Muller:1977mm}, 
and to simplify, we consider a reference frame 
such that either only an electrical field is present ($B=0$),
 or only a magnetic field is present ($E=0$).  This requires 
the invariant $E\cdot B$ vanish identically, as can easily 
be arranged for laser-driven experiments. However, our remarks are
more generally valid for all field configurations.  For these
two  cases,  one can write~\cite{Labun:2008qq}:
\begin{equation}\label{Ef}
{\cal L}_{\rm eff}(E) = 
\frac{m^2}{8\pi^2\beta }\int_0^{\infty}\!\!\!d\nu \,
       \ln\left(\frac{\nu^2-m^2+i\epsilon}{m^2}\right)\ln(1-e^{-\beta \nu}),
\end{equation}
\begin{equation}\label{Bf}
{\cal L}_{\rm eff}(B) = 
\frac{m^2}{8\pi^2\beta }\int_0^{\infty}\!\!\!d\nu\,
       \ln\left(\frac{\nu^2+m^2}{m^2}\right)\ln(1-e^{-\beta \nu}),\\
\end{equation}
where  $\beta = m\pi /eE$ or  $\beta = m\pi/eB$, respectively.  
For comparison, the functional in statistical physics corresponding 
to the effective action, the free energy, is for Bosons and Fermions 
(upper and lower signs, respectively)
\begin{equation}
{\cal F}_{\rm B/F}=\pm\frac{1}{\beta}\sum_k\ln(1\mp e^{-\beta \nu_k}).
\end{equation}
The sum is over all modes and it is common to transit from discrete 
to continuous sum in which case it is understood that one normalizes 
dividing by the volume $V$.  ${\cal F}$ is then the free energy density just 
as ${\cal L}$ is the (effective) action density.

We see that beginning from a microscopic theory of fermion pair 
vacuum fluctuations, the effective EHS action acquires a form typical 
for bosons in a thermal bath.  Confirmation of the statistics (sign) 
reversal comes from inspection of the spin-0 effective action for 
which the same form is found but with 
$\ln(1-e^{-\beta M})\to (1/2)\ln(1+e^{-\beta M})$, where the 
factor 2 accounts for reduced number of degrees of  freedom. 

Another difficulty is that the temperature $\beta^{-1}$ 
is half of the value that would  correspond to the Hawking-Unruh effect: 
\begin{equation}
T_{\rm HU}=\frac {a}{2\pi}; \qquad a=\frac{qE}{m}; \qquad  \beta_{\rm HU}
         =\frac{1}{T_{\rm HU}}=\frac{2\pi m}{qE}=2 \beta.
\end{equation}
There is at present no understanding of why the sign reversal 
occurs, nor why the temperature differs from the only comparable 
quantity by a factor two.  These results appear also in the 
classical WKB limit of quantum field theory and in many 
related methods.  Further discussion is found in 
Pauchy Hwang and Kim~\cite{PauchyHwang:2009rz}.  

An interpretation of vacuum fluctuations in the presence 
of electric field as if there were a thermal bath is inviting 
yet not quite consistent with the interpretation of the 
accelerated frame observing a thermal bath as in the picture 
arising from the Hawking-Unruh effect.  This disagreement 
implies that General Relativity, the Equivalence Principle, 
Quantum Field Theory and Quantum Electrodynamics remain 
startlingly inconsistent despite 50+ years of effort.  

The mutually inconsistent  temperature interpretations arising from 
both external fields and acceleration evoke another interesting 
and challenging problem clearly related, but somewhat to the 
side of our current discussion.   When electrical fields decay 
into a multitude of particle-anti-particle pairs,  a coherent, 
pure quantum state materializes into a multi-particle high 
entropy state. There is no heat bath, no coupling to an environment. 
 Yet rapidly a lot of entropy is created as if entropy 
also materialized, or the system had access to a hidden 
entropy source.  

This problem is already 
intensively studied considering rapid entropy production 
at times of formation of quark-gluon plasma but 
remains unresolved. One must be aware that sudden
appearance of entropy in an isolated system 
associated with a former pure quantum state 
violates the principles of quantum mechanics. We note that this 
effect only arises in presence of critical acceleration when
massive particle production ensues. Thus  
 we learn that also quantum mechanics may need modification
when critical acceleration is reached.  Though our prior
discussion of non-Machian physics addressed mainly forces, the
entropy crisis places quantum mechanics also among 
theories in need of Machian extension.  

 Nevertheless, 
the difficulties of QED underline our major insight: 
 the classical theory of electromagnetism (both classical 
and quantum) is incomplete and fails in the strong acceleration 
limit.  Considering the equivalence principle and the Machian 
nature of Einstein's gravity we know that an extension of the 
theory of electromagnetism to the domain of strong acceleration 
will need to be consistent with geometric gravity and Mach3. 
Because the problem exists in full force in deeply classical 
domain, it is not necessary to develop quantum gravity in 
order to understand electromagnetism in the 
strong acceleration limit. Any improvement in the action
of charged matter and EM fields will trickle down both to 
quantum theory and quantum field theory.

%%%%%%%%%%%%%%%%%%%%%%%%%%%%%%%%%%%%%%%%%%%%%%%%%%%%%%%%%%%%%%%%%%%%%%%%%%%%%%%%
\begin{theacknowledgments}
We would like to thank B.M. Hegelich  (Los Alamos), H. Ruhl and P. Thirolf (LMU-Munich) for 
illuminating discussions, and Herrn Dietrich Habs for his interest and  enthusiastic 
Bavarian hospitality at the Munich Centre of Advanced Photonics -- DFG excellence 
cluster in Munich. 
This work was supported by the DFG Cluster of Excellence 
MAP (Munich Centre of Advanced Photonics), and by a grant from: the
U.S. Department of Energy  DE-FG02-04ER41318.
\end{theacknowledgments}

\end{document}